\documentstyle[12pt]{article}
\newcommand{\be}{\begin{equation}}
\newcommand{\ee}{\end{equation}}
\newcommand{\lb}{\label}
\newcommand{\rf}[1]{(\ref{#1})}
\newcommand{\cao}{\c c\~ao\ }

\newcommand{\half}{\frac{1}{2}}
\newcommand{\thalf}{\mbox{\small $\frac{1}{2}$\normalsize}}
\newcommand{\ttt}{\mbox{\small $\frac{3}{2}$\normalsize}}
\newcommand{\tsqrt}{\mbox{\small $\frac{1}{\sqrt{2}}$\normalsize}}
\begin{document}
\title{Critical Behaviour of Mixed Heisenberg Chains}
\author{ F. C. Alcaraz\thanks{e-mail: alcaraz@power.ufscar.br}, A. L. Malvezzi\thanks{e-mail: palm@power.ufscar.br} \\
	Departamento de F\'\i sica \\ Universidade Federal de S\~ao Carlos \\
	 13565-905, S\~ao Carlos, SP, Brazil}
\date{}
\maketitle
\begin{abstract}
\vspace{0.2cm}
The critical behaviour of anisotropic Heisenberg models with two kinds of 
antiferromagnetically exchange-coupled centers are studied numerically by
using finite-size calculations and conformal invariance. These models exhibit
the interesting property of ferrimagnetism instead of antiferromagnetism.
Most of our results are centered in the mixed Heisenberg chain where we
have at even (odd) sites a spin-$S\ (S')$ SU(2) operator interacting with a
XXZ like interaction (anisotropy $\Delta$). Our results indicate universal
properties for all these chains. The whole phase, $1>\Delta>-1$, where
the models change from ferromagnetic $( \Delta=1 )$ to 
ferrimagnetic $(\Delta=-1)$ behaviour is critical. 
Along this phase the critical fluctuations are 
ruled by a $c=1$ conformal field theory of Gaussian type. 
The conformal dimensions 
and critical exponents, along this phase, are calculated by studying these 
models with several boundary conditions.
\end{abstract}
\section{Introduction}

The critical properties of one-dimensional regular Heisenberg spin chains
with one kind of antiferromagnetically exchange-coupled spins have been
extensively studied in the literature. The prototype of these models is the
anisotropic $S=\half$ 
Heisenberg model or XXZ chain\cite{1}. This model is exactly 
integrable with a critical line of continuously varying critical exponents
as we change the anisotropy $(\Delta)$, bringing the model from the
ferromagnetic $(\Delta=1)$ to the antiferromagnetic $(\Delta=-1)$
isotropic points. With the advance of the conformal invariance ideas\cite{2} the
whole operator content of this model was obtained\cite{3,4}. 
The critical fluctuations are
governed by a Gaussian type conformal field theory with conformal anomaly
$c=1$ and, moreover, the underlying currents satisfying a U(1) Kac-Moody
algebra\cite{5}.

The extension of the XXZ chain to higher spins $S>\half$ attracted 
considerable attention after Haldane\cite{6} conjectured that, for the 
isotropic antiferromagnetic point
$(\Delta=-1)$, the model is critical or not depending if $S$ is
half-odd-integer or integer, respectively. Consistent with this conjecture,
numerical calculations\cite{7,8,9} indicate 
that in the case of half-odd-integer spins the
models are critical in the whole range of anisotropies
 $(1\geq\Delta\geq-1)$ from the ferromagnetic to the antiferromagnetic
point. In the case where $S$ is integer, a critical line starting at the
ferromagnetic point ends at $\Delta_c$ 
before the antiferromagnetic phase $(\Delta_c >-1)$
entering the massive Haldane phase\cite{7,8,9}. 
For all spins the massless phases are
ruled by a $c=1$ Gaussian-like conformal field theory\cite{8}.

In this paper we extend these studies by studying a 
 quantum chain in which two types of antiferromagnetically exchange-coupled 
spins $S$ and $S'$ are located at alternate sites. When $S\neq S'$, according
to a theorem due to Lieb and Mattis\cite{10}, 
the isotropic model $(\Delta=-1)$ exhibits 
ferrimagnetic order, with a $(S-S')\frac{L}{2}$-degenerate ground state,
where $L$ is the chain length. Consequently as we vary the anisotropy the 
model goes from the ferromagnetic point $(\Delta=1)$ to the ferrimagnetic 
point $(\Delta=-1)$. We studied these models for $(S,S')=(\half,1)$
 and $(S,S')=(\half,\frac{3}{2})$, using finite-size scaling and conformal 
invariance\cite{2}.
To supplement our studies we considered two other anisotropic models that
also exhibit ferrimagnetic behaviour at the isotropic point.
Our studies show that all these models between the two 
(ferromagnetic and ferrimagnetic) isotropic points $1>\Delta>-1$ 
have a universal Gaussian critical 
behaviour with central charge $c=1$. In this massless phase the critical
exponents exhibit a model-dependent variation with the anisotropy.

The paper is organized as follows. In section 2 we define 
the \linebreak $(S,S')-$Heisenberg chain and review some results obtained from 
conformal invariance in the model with $S=S'$. In section 3 and 4 we 
present our numerical results for the $(S,S')-$Heisenberg model and
related models. Finally in section 5 we present our general conclusions.

\section{The model and conformal invariance relations}

The mixed Heisenberg quantum chains are defined by attaching an SU(2) 
spin-$S$ at the odd sites 
$( \vec{\sigma}_i = ( \sigma^x_i,\sigma^y_i,\sigma^z_i );\ i = 1,3,5, \ldots )$
and a spin $S'$ at the even sites 
$( \vec{S}_i = ( S^x_i,S^y_i,S^z_i );\ i = 2,4,6, \ldots )$. The Hamiltonian
on an $L$ (even) site chain, with periodic ends, is defined by
\be
\lb{H}
H = -\,\sum_{i=1}^{L/2}\,\left(\,\sigma_{2i-1}^{x}S_{2i}^{x} +  
\sigma_{2i-1}^{y}S_{2i}^{y} + \Delta\sigma_{2i-1}^{z}S_{2i}^{z}\,\right)
\ee
where $\Delta$ is the anisotropy constant. This Hamiltonian, like the
standard spin-$S$ XXZ chain $(S=S')$, has a U(1) symmetry due to its 
commutation with the $z$ component of the total spin
\be
\lb{U1}
S_z = \sum_{i=1}^{L/2}\,\left(\,\sigma_{2i-1}^z + S_{2i}^z\,\right)
\ee
For $\Delta>1$ the model is massive and ferromagnetic with a double
degenerate ground state corresponding to the two fully ordered states with
$S_z = \pm\frac{L}{2}(\,S + S'\,)$. 
At $\Delta=1$ the lowest energy in all U(1)
 sectors \linebreak $\left(\,S_z = -\frac{L}{2}(\,S + S'\,),-\frac{L}{2}(\,S + S'\,)+1,
\ldots ,\frac{L}{2}(\,S + S'\,)-1,
\frac{L}{2}(\,S + S'\,)\,\right)$ are 
degenerate, rendering a ferromagnet ground state with total spin 
$\frac{L}{2}(\,S + S'\,)$ and a massless spectra with a quadratic dispersion 
relation.
For $1>\Delta>-1$ the ground state is single or double degenerate, depending 
if $| S - S' |\frac{L}{2}$ is integer or half-odd-integer and belongs to the
sectors with $S_z = 0$ or $S_z=\pm \frac{1}{2}$, respectively. At $\Delta=-1$
the lowest energies in the sectors where $S_z = -\frac{L}{2}|S-S'|,
-\frac{L}{2}|S-S'|+1, \ldots , \frac{L}{2}|S-S'|-1,\frac{L}{2}|S-S'|$
become degenerate and we have ferrimagnetic order\cite{10}. 
For $\Delta < -1$ 
the ground state is double degenerate, occuring in the sectors with
$S_z = \pm \frac{L}{2}|S-S'|$, and we expect a massive behaviour as in the
standard $S=S'$ XXZ chain. 
In order to illustrate the spectral dependence on 
the anisotropy $\Delta$, in Fig. 1 we draw in schematic form the location of
the lowest eigenenergies of \rf{H} in the various $S_z$ sectors.

Our analysis indicates that in the whole region $1\geq\Delta\geq-1$ the model
is critical, like the $S=\half$ XXZ model $(S=S'=\half)$. We assume that the
Hamiltonian \rf{H}, like most statistical mechanics quantum
 chains are conformally invariant in its critical regime. Under this 
 assumption the machinery arising from conformal invariance tells us that for
 each primary operator\cite{2,11} 
$O_\alpha$ with dimension $x_\alpha$ and spin $s_\alpha$ 
in the Virasoro operator algebra of the infinite system, there exists an
 infinite tower of states, in the quantum Hamiltonian, for a periodic chain
 of $L$ sites, whose energy and momentum as $L \rightarrow \infty$ are given by
\be
\lb{Etower}
E_{j,j'}^{\alpha} = E_{0}(L) + \frac{2\pi}{L}v( x_\alpha + j + j' ) + 
o\left( L^{-1} \right)
\ee
and
\be
\lb{Ptower}
P_{j,j'}^{\alpha} = \frac{2\pi}{L}(\, s_\alpha + j - j' )
\ee
where $j,j' = 0,1, \ldots$. Here $E_{0}(L)$ is the ground-state energy and $v$
is the velocity of sound, 
which can be determined by the energy-momentum dispersion
 relation or from the difference among consecutive energy levels in a same
conformal tower. The finite-size corrections of the ground-state energy also
give a way to calculate the conformal anomaly. For periodic chains, the
 ground-state energy behaves asymptotically as\cite{12}
\be
\lb{cc}
\frac{E_{0}(L)}{L} = e_\infty - \frac{\pi c v}{6 L^2} +
 o\left( L^{-2} \right)
\ee
where $e_\infty$ is the ground-state energy per site in the bulk limit.

In the case where $S=S'$ a critical phase appears\cite{7,8,9} 
in \rf{H} for anisotropies 
$1\geq\Delta\geq\Delta_{c}(S)$, where due to the Haldane conjecture\cite{6},
$\Delta_{c}=-1$ or $\Delta_{c}>-1$ depending if $S$ is half-odd-integer or
not. This massless phase is described by a U(1) conformal field theory with
 central charge $c=1$\cite{8}. 
The  
anomalous dimensions $x_{\alpha}$ appearing through \rf{Etower} 
in the U(1) sector $S_z = n$ of the Hamiltonian 
\rf{H} with periodic ends are given by
\be
\lb{xnm}
x_{n,m} = n^{2}x_p + \frac{m^2}{4x_p} \ \ \ , \ \ \ n,m = 0,\pm 1,\pm 2, \ldots
\ee
where $x_p$ depend on $S$ and $\Delta$. 
For $S=S'=\half$ we have the exact 
dependence\cite{3,4} $x_p = \left(\pi - \cos^{-1}(-\Delta)\right)/2\pi$. 
Although the model is not integrable for 
$S=S'>\half$, numerical calculations indicate the conjecture\cite{8} \\ 
$x_p = \left(\pi - \cos^{-1}(-\Delta)\right)/4\pi S$
 for $-1<\Delta\put(5,2.5){$<$}\put(5,-3.3){$\sim$}\hspace{.75cm}0$. 
Beyond the dimensions \rf{xnm} other integer dimensions 
also appear in the sector with $S_z = n = 0$. This fact indicate that 
the underlying conformal field theory satisfies a larger algebra than the 
Virasoso conformal algebra, namely, a U(1) Kac-Moody algebra\cite{5,8}.
 The dimensions 
\rf{xnm} correspond to operators $O_{n,m}$ and the number of its descendants
 will be given by the product of two U(1) Kac-Moody characters. The dimensions
\rf{xnm} indicate that the operators $O_{n,m}$ correspond to the Gaussian model 
operators\cite{13} composed of a spin-wave excitation of index $n$ and a
``vortex" excitation of vorticity $m$.

Other interesting properties of these $c=1$ critical phases appear when we 
consider these chains with more general boundary conditions, compatible 
 with its U(1) symmetry, {\em i.e.}, by preserving the total spin $S_z$ as
a good quantum number. Two of these conditions are, the $x-y$ twisted boundary 
conditions
\be
\lb{twis1bc}
S_{L+1}^{x} \pm iS_{L+1}^{y} = 
e^{\pm i\Phi}\,\left( S_{1}^{x} \pm iS_{1}^{y} \right)
\ee
\be
\lb{twis2bc}
S_{L+1}^{z} = S_{1}^{z},
\ee
where $\Phi$ is an arbitrary angle, and free boundary conditions
\be
\lb{freebc}
S_{L+1}^{x} = S_{L+1}^{y} = S_{L+1}^{z} = 0.
\ee

The net effect of the boundary angle $\Phi$ in the dimensions \rf{xnm} is to 
shift the spin-wave index by an amount\cite{3,8} $\Phi/2\pi$,
\be
\lb{xnmPhi}
x_{n,m+\Phi/2\pi} = n^{2}x_p + \frac{(m+\Phi/2\pi)^2}{4x_p}.
\ee
In a semi-infinite lattice the correlation functions involving lattice 
points near the surface have a power-law decay distinct from the case 
where the points are away from the surface (bulk behaviour). These 
correlations are ruled by the surface exponents $x_s$. These exponents 
can  be obtained from the finite-size corrections of the mass-gap 
amplitudes of finite chains with free ends. 
  Instead of \rf{Etower} and \rf{Ptower}, to each surface 
exponent of the semi-infinite system, at the critical point, 
there exists a set of states with energies 
given by\cite{11}
\be
\lb{EtowerFree}
E_{r}^{(F)} = E_{0}^{(F)}(L) + \frac{\pi v}{L}( x_s + r ) + o\left( L^{-1} \right)
\ee
where $E_{0}^{(F)}(L)$ is the ground-state energy of the $L$-site chain
 and $r = 0,1,2, \ldots$. 
Instead of  \rf{cc}  we have\cite{12}
\be
\lb{ccFree}
\frac{E_{0}^{(F)}(L)}{L} = e_\infty + \frac{f_{\infty}}{L} 
- \frac{\pi c v}{24 L^2}  + o\left( L^{-2} \right)
\ee
where $f_\infty$ is the bulk limit of the surface energy. The study of 
\rf{H} with $S=S'$ and free ends\cite{8,14} shows that in the critical region, for
each sector $S_z = n$ there appears only one conformal tower associated to the 
dimensions
\be
\lb{xs}
x_{s}(n) = 2n^2 x_p \ \ \ , \ \ \ n = 0,1,2, \ldots
\ee
with the multiplicity of its descendants given by the character of a 
single U(1) Kac-Moody algebra\cite{8,15}.

\section{Results for the mixed Heisenberg chains}

We calculate numerically the eigenspectra of the Hamiltonian \rf{H} by using
the Lanczos method in the cases where $(S,S')=(\half,1)$ and $(S,S')=(\half,\frac{3}{2})$
up to lattice sizes $L=20$ and $L=16$, respectively. Our results, for several 
boundary conditions, indicate that in the whole region $1\geq\Delta\geq-1$
the model is gapless and conformally invariant. The ground state in this region
 will have the lowest possible value of $|S_z|$. It is non degenerate
 with $S_z = 0$ if $\frac{L}{2}|S-S'|$ is integer and is doubled degenerate 
with $S_z = \pm \half$, otherwise. Consequently in order to obtain a uniform 
convergence of our finite-size results we consider in the case 
$(S,S')=(\half,1)$ only lattice sizes which are multiples of 4.

Let us consider initially the case of periodic chains. The model is invariant
under translation by a unit cell with two spins, with momentum 
$p = \frac{4\pi}{L} l \ \ ( l = 0,1, \ldots , \frac{L}{2}-1 )$. In order to
calculate the conformal anomaly and exponents from (\ref{Etower}-\ref{cc})
we should estimate the sound velocity. As in the $S=S'=\half$ case the lowest
eigenenergy with nonzero momentum ( modulo $\pi$ ) belonging to the 
 ground-state sector is associated to a primary spin-1 operator with dimension
equal to unity for all values of $\Delta$. Using Eq. \rf{Etower} we
obtain an estimate for the sound velocity
\be
\lb{zeta}
v(L) = \frac{\left( E_{4\pi/L} - E_{0}(L) \right) L}{2\pi}
\ee
where $E_{0}(L)$ is the ground-state energy and $E_{4\pi/L}$ is the lowest
eigenenergy of a state with momentum $4\pi/L \ ( mod\, \pi )$. Using
\rf{zeta} the conformal anomaly $c$ is obtained by extrapolating the numerical 
sequence obtained from  \rf{cc}. In table 1 we show, for some values of 
$\Delta$, our estimates for $c$ in the two cases 
$(S,S')=(\half,1)$ and $(S,S')=(\half,\frac{3}{2})$. 
All the extrapolated results
 reported in this table, as in the subsequent ones, are calculated by
using the alternating $\epsilon$-algorithm\cite{16}, which is a variant of the
 van den Broeck-Schwartz method\cite{17}. The errors are roughly estimated from the
 region of stability of these approximants.  It was not possible to obtain
 reliable results near the isotropic points $\Delta=1$ and $\Delta=-1$. 
This happens because like near the ferromagnetic models with $S=S'$\cite{8}, the
sound velocity decreases towards zero as we tend towards the isotropic point
(the energy-momentum dispersion relation changes from
linear to quadratic). Our results indicate that we have a conformal anomaly
 $c=1$ in both cases and we believe that this is the general case for
arbitrary $S \neq S'$, since the spectrum (see figure 1) shows the same 
essential features independently of $S$ and $S'$ being integer or
half-odd-integer. Moreover the vanishing of the sound velocity as we tend
 toward the ferromagnetic $( \Delta=1 )$ and ferrimagnetic $( \Delta=-1 )$ 
points  indicate that the critical fluctuations around the ferrimagnetic
point are similar to those near the ferromagnetic point. As we see in  
table 1 the results for $(S,S')=(\half,\frac{3}{2})$ are slightly better than those
 of $(S,S')=(\half,1)$.
 This is due to the fact that for $(S,S')=(\half,\frac{3}{2})$
 the number of terms in the finite-size 
sequences are larger since $L$ can be an arbitrary even number.

The conformal dimensions are calculated by using  \rf{Etower} and
 \rf{zeta}. For example in the $(S,S')$ chain the lowest energy $E_n$ in
the sector with total spin $S_z = n$ is associated to the dimension 
$x_{n}^{S,S'}$, which is calculated from the asymptotic $(L\rightarrow\infty)$
 value of the sequence
\be
\lb{xn1}
x_{n}^{S,S'}(L) = \frac{E_{n}(L) - E_{0}(L)}{2\pi v(L)}
\ee
where $v(L)$ is given by \rf{zeta}. In tables 2 and 3 we show our results
for $n=1$ and 2 and some values of the anisotropy. We see from these tables
 that for all values of the anisotropy, the extended relation 
\be
\lb{xn2}
x_{n}^{S,S'} = n^{2} x_{1}^{S,S'}
\ee
holds. These dimensions are similar to the Gaussian dimensions $x_{n,0}$
appearing in \rf{xnm}, on identifying $x_{p} = x_{1}^{S,S'}$. Different
from  the critical regime in the homogeneous spin case $S=S'$, when $S \neq S'$
 the dimensions $x_{1}^{S,S'}$ 
increase as we depart from the ferromagnetic point, but 
around $\Delta \put(5,2.5){$<$}\put(5,-3.3){$\sim$}\hspace{.75cm}  -0.5$ 
it starts to decrease again and we have small values of $x_p$  
near the ferrimagnetic point, as we normally see near the
ferromagnetic point. The small values of the exponents $x_{1}^{S,S'}$ near
the ferromagnetic and ferrimagnetic points is the signature of the long-range
 ordered ground state at the isotropic points. The fourth column of these
 tables also shows that near the ferromagnetic point 
$x_{1}^{S,S'} = x_{1}^{1/2,1/2}/(S+S')$, which give us
\be
\lb{x1}
x_{1}^{S,S'} = \frac{\pi - \cos(-\Delta)}{2\pi(S+S')} \ \ \ , \ \ \ 
\Delta \rightarrow 1 \ .
\ee

This result when compared with the conjectured\cite{8}
 results for $S=S'$, indicate
 that near the ferromagnetic point we have essentially a Heisenberg model with
 effective spin $(S+S')/2$. On the other hand the degenerascy of the 
ground state at the ferrimagnetic point $(\Delta=-1)$ induce us to expect near
 this point an effective Heisenberg ferromagnetic chain with effective spin
 $(S-S')/2$. However, we are not able to make a conjecture like in Eq. \rf{x1}.

We also made an independent calculation of the exponents $x_{1}^{1/2,1}$ by
 using standard finite-size scaling\cite{18}. This exponent is related to the 
ratio of the ``electric" susceptibility $\gamma_p$\cite{19} 
and correlation-lenght exponent
$\nu$, by the relation
\be
\lb{gon}
\frac{\gamma_p}{\nu} = 2( 1 - x_{1}^{1/2,1} ).
\ee
The ``electric" susceptibility is the response of the system to an 
staggered transversal field. This susceptibility $\chi_L^{\xi}$ is calculated  by adding an  
``electric field" interaction 
$\xi \sum_{i} \left( g_{A}\sigma_{i}^{x} + g_{B}S_{i+1}^{x} \right)$
in \rf{H} 
\be
\lb{susc}
\chi_{L}^{\xi} = 
\left. \frac{\partial^2 E_{0}(\xi)}{\partial \xi^2} \right|_{\xi=0}
\ee
where $E_{0}(\xi)$ is the ground-state energy in the presence of the
``electric" field. The Land\'e factors $g_A$ and $g_B$ produces the 
staggering effect of the transverse "electric" field. In the isotropic 
case $S = S'$ 
we must choose  $g_A/g_B \neq 1$, otherwise it will produce the effect 
of a uniform transverse field, which is not related with the 
exponent $\gamma_p$. 
In the case $S \neq S'$ our results shows that a similar effect also 
occur and in order to calculate $\gamma_p$ we should consider $g_A/g_B \neq 
r_c$. 
  For $S=1/2$ and $ S'=1$,  $r_c$ changes  from 
$r_c = 2$ to $r_c = 2.66$ as the anisotropy changes from the ferromagnetic
point $(\Delta=1)$ to the ferrimagnetic one $(\Delta=-1)$. These points are
probably related to the compensation mechanism which usually happens in 
ferrimagnetic ordered models when temperature effects are taken into
account\cite{20}. 
In the bulk limit $\chi_{L}^{\xi} \sim L^{\gamma_{p}/\nu}$. 
%
%By extrapolating the sequences obtained from \rf{gon} with $L$ up to 16 we
%obtain the results shown in the last column of table 2, with a good 
%agreement with those derived by using the conformal invariance relations. 
%
The extrapolation of the finite-size sequence obtained by choosing
$g_A = 0$ and $g_B = 1$ gives from \rf{gon} the results in the last
column of table 2. In this case, since we calculate lattices up to $L=16$,
it is difficult to obtain an error estimate through the alternating 
$\epsilon$-algorithm\cite{16}. 
The results presented are in reasonable agreement with
those derived by using the conformal invariance relations.
Beyond the dimensions presented in tables 2 and 3 our results also 
indicate other dimensions which would correspond to $x_{n,m}$ in \rf{xnm} 
with $m \neq 0$.
Instead of presenting these dimensions we show in table 4 the lowest dimensions
$x_{\Phi}^{S,S'}$ obtained by calculating the $(S,S')$ model with the twisted
boundary conditions given by  (\ref{twis1bc},\ref{twis2bc}). These 
dimensions are obtained from the bulk limit extrapolations of the sequence
\be
\lb{xPhiL}
x_{\Phi}^{S,S'}(L) = \frac{E_{\Phi}(L) - E_{0}(L)}{2\pi v(L)}
\ee
where $E_{\Phi}(L)$ is the ground-state energy of the Hamiltonian \rf{H}
with $L$ sites and boundary conditions (\ref{twis1bc},\ref{twis2bc}). The
results given in table 4 indicate that
\be
\lb{xPhi}
x_{\Phi}^{S,S'} = \frac{(\Phi/2\pi)^2}{4x_{1}^{S,S'}}
\ee
where $x_{1}^{S,S'}$ is given in tables 2 and 3. These dimensions correspond to
 the dimensions $x_{0,\Phi/2\pi}$ in \rf{xnmPhi}. The results presented in
tables 2-4 clearly indicate that in the whole disordered regime $-1<\Delta<1$
the conformal dimensions are those of a Gaussian model. The dimensions are 
given by \rf{xnm}  where $x_p = x_{1}^{S,S'}$ is a continuous function of 
$\Delta$ with some of its values given in tables 2 and 3. We have also studied
 the $(S,S') = (\half,1)$ model with lattice sizes $L = 4l + 2\ \ (l=1-4)$.
In this case the ground state is degenerate like in the standard 
$S=S'=\half$ XXZ chain with an odd number of sites\cite{2} and we obtain the 
dimensions $x_{n+1/2,m}\ \ (n,m=0,\pm 1,\pm 2, \ldots)$.

For completeness, we also calculated the surface exponents of these chains. 
These exponents are calculated from the eigenspectra of \rf{H} with free
 boundary conditons. From \rf{EtowerFree} the surface exponents 
$x_{s}^{S,S'}(n)$ associated with the state with lowest energy $E_{n,0}^{(F)}$ 
in the sector $S_z = n$ of the $(S,S')$ chain can be estimated from the 
large-$L$ behaviour of the sequence
\be
\lb{xFree}
F^{S,S'}(n,L) = 
\frac{E_{n,0}^{(F)} - E_{0,0}^{(F)}}{E_{0,1}^{(F)} - E_{0,0}^{(F)}}
\ee
where $E_{n,m}^{(F)}$ is the $m$ excited state in the sector $S_z = n$.
The estimator \rf{xFree} was obtained by assuming that like in the $S=S'$ 
case, the first mass gap amplitude in the ground-state sector is associated
with a dimension equal to unity. Our results are shown in table 5 where
we clearly see the same type of extended relation as in \rf{xn2}, namely
\be
\lb{xnS}
x_{s}^{S,S'}(n) = n^{2} x_{s}^{S,S'}(1).
\ee
Comparing these results with those of  tables 2 and 3 we obtain the relations
\rf{xs} expected in a Gaussian model
\be
\lb{xsSS'}
x_{s}^{S,S'}(n) = 2n^2 x_{1}^{S,S'} \ \ \ , \ \ \ n = 0,1,2, \ldots
\ee
where $x_{1}^{S,S'}$ is the dimension which appeared in the periodic case.

\section{Results for other related models of ferrimagnetism}

Inspired by the results of last section we will try to see if the general
critical features of the $(S,S')$-Heisenberg models can also be observed in
other models exhibiting ferrimagnetism. In this direction we will study two
 other models defined on a bipartite lattice as in figures 2a and 2b. At each
 lattice point we attach a spin-$\half$ operator which interacts along the lines
 of Fig. 2 with interactions of XXZ type
\be
\lb{H2}
H = -\,\sum_{<ik>}^{L}\,\left(\,\sigma_{i}^{x}\sigma_{k}^{x} +  
\sigma_{i}^{y}\sigma_{k}^{y} + \Delta\sigma_{i}^{z}\sigma_{k}^{z}\,\right).
\ee
The first model defined in the bipartite lattice of figure 2a we denote
$ABC$ and the second one in figure 2b we denote $AB_2$. The lattice size
$L$ is considered as two times the number of lattice sites in the sublattice
 $A$ (in figures 2a and 2b $L=6$). Both models, at the isotropic
 ferromagnetic point $\Delta=1$ are fully ordered. At $\Delta=-1$ they
show a ferrimagnetic behaviour since the total number of spin variables
 in each sublattice is not equal. As before the Hamiltonian has a U(1)
 symmetry and its Hilbert space is separated in the $\sigma^z$-basis into 
block disjoint sectors labelled by the $z$-component of the total spin
$S_z = \sum_{i} \sigma_{i}^z$. The ground state location in these sectors as
well as its degeneracies on a finite lattice for $-1\leq\Delta\leq1$ are
exactly like those shown in figure 1, on taking $S=\half$ and $S'=1$.

Our numerical results for periodic boundary conditions indicate that both
models are disordered and massless in the whole regime $1\geq\Delta\geq-1$.
The thermal effects of the model $AB_2$ at $\Delta =-1$ are considered in 
\cite{21}. We now report our results separately for both models.

\subsection{Model $ABC$}

Using equation \rf{zeta} we observe as in section 3 that the sound velocity 
approaches zero as we tend towards the isotropic points $\Delta=\pm 1$. The
conformal dimensions $x_{n}^{ABC} \ \ (n=1,2,\ldots)$ associated to the lowest
eigenenergy in the sector $S_z = n$ are obtained by extrapolating the sequence
 \rf{xn1} for $L$ up to 16. Our results for some dimensions and anisotropies
are shown in table 6. As we see in this table, like in \rf{xn2} the
relation $x_{n}^{ABC} = n^2 x_{1}^{ABC}$ also holds for the whole range of
anisotropies $1\geq\Delta\geq-1$, with $x_{1}^{ABC}$ depending continuously on
$\Delta$. Since at $\Delta=1$ we have the 
same ground-state degeneracy as in the
$(S,S')=(\half,1)$ Heisenberg model we would expect an asymptotic behaviour, as
$\Delta \rightarrow 1$, like \rf{x1} with $S=\half$ and $S'=1$. However the
fourth column of table 6 tell us this is not true. We have also studied this
model with the twisted boundary conditions \rf{twis1bc} and \rf{twis2bc}
with results as predicted in \rf{xnmPhi}, which clearly indicate an 
underlying $c=1$ Gaussian field theory in the whole regime 
$1\geq\Delta\geq-1$.

\subsection{Model $AB_2$}

In this case, beyond the U(1) symmetry, we also have a Z(2) local gauge 
invariance corresponding to an independent interchange of the spin variables
located at points like those shown in broken-lined rectangles 
in figure 2b. Since we have $L/2$ disjoint sectors labeled by the eigenvalues 
$g_l \ \ (\pm 1)$ of the gauge operators
\be
\lb{Gl}
G_{l} = \vec{\sigma}_{i}.\vec{\sigma}_{k} + \frac{1}{4} \ \ \ ; \ \ \ 
l = 1,2, \ldots ,L/2 ,
\ee
where $\vec{\sigma}_{i}$ and $\vec{\sigma}_{k}$ are the spin-$\half$ operators 
located at the $l^{\rm th}$ rectangle,
a sector having $g_l = 1$ will be spanned in a basis with three
even combinations of the spin variables
 $\vec{\sigma}_{i}$ and $\vec{\sigma}_{k}$
located at the retangle $l$. This means that in $\sigma^z$-basis we should 
have the triplet combination $|++\rangle,\tsqrt ( |+-\rangle + |-+\rangle )$
and $|--\rangle$. On the other hand if $g_l = -1$ we should have a singlet
 combination $\tsqrt ( |+-\rangle - |-+\rangle )$.
It is not difficult to verify that the interaction between spins in the
sublattice $A$  with a given neighbouring retangle $l$ with $g_l = 1$
 (triplet) is exactly the same as in the $(S,S') = (\half,1)$ Heisenberg
interaction ( see \rf{H} ). In contrast, the interaction with
a retangle with $g_l = -1$ (singlet) is zero. This implies some interesting
consequences. For a given U(1) sector $S_z = n$ the eigenenergies of the 
gauge sector with $g_l = 1$ for all $l=1,2, \ldots ,L/2$ of the Hamiltonian
 $AB_2$ with periodic ends will be exactly the same as the $S_z = n$ sector
 of the $(S,S')=(\half,1)$ Heisenberg chain \rf{H}, also with periodic boundary
 condition. For general gauge choices we lose translation invariance since
the operator \rf{Gl} does not have this symmetry. However, for the gauge choices
$g_{1}=g_{2}= \ldots =g_{L/2} = \pm 1$, this invariance is recovered and
we obtain the same dimensions as in the $(S,S') = (\half,1)$ Heisenberg model
studied in section 3.

The eigenenergies in the sectors with $g_l = -1$ of the $AB_2$ model with
periodic ends will correspond to the composition of energies of the
$(\half,1)$-Heisenberg chains with 
free boundary conditions and different lattice
sizes. This exact correspondence together with the relation \rf{ccFree}
imply that the lowest energy in these sectors, in the bulk limit, will have a
finite gap when compared with the ground-state energy, which happens in the
sector $g_{1}=g_{2}= \ldots =g_{L/2} = 1$. This gap is proportional to the
surface energy $f_\infty$ of the related $(S,S')=(\half,1)$ Heisenberg chain.
This produces the interesting feature that the correlation functions of 
operators
which commute with \rf{Gl} will have a power-law decay with exponents like
those of the periodic $(\half,1)$-Heisenberg chain, while correlations of 
non-commuting operators may exhibit an exponential decay, with rate proportional
to $f_\infty$. An example of such operators is 
$\sigma_{k}^{z}\sigma_{k}^{+}\sigma_{l}^{-}$ where $k$ and $l$ are indices 
inside a given retangle in Fig. 2b and 
$\sigma^{\pm} = \sigma^{x} \pm i\sigma^{y}$. Apart from these pathological
correlations most of them will be of the same nature as those of the 
$(S,S')=(\half,1)$ Heisenberg model and our results of section 3 indicate that
they are described by a Gaussian like field theory in the regime 
$1>\Delta>-1$.

\section{Conclusions}

Anisotropic quantum chains with one kind of spin $S$ exhibit a critical phase
 with continuously varying exponents governed by a $c=1$ Gaussian like
conformal field theory. This phase starts at the ferromagnetic point $\Delta=1$
 with an endpoint at $\Delta=\Delta_{c}(S)$, which for half-odd-integer $S$ is
expected to be 1 and $\Delta_{c}(S)>-1$ otherwise. This means that the 
antiferromagnetic point has quite different physics depending on the parity
of $2S$. In this paper we analyse anisotropic Heisenberg chains with two kind
of exchange-coupled centers. Due to a noncompensation effect, these models
show ferrimagnetism instead of antiferromagnetism. As we change the anisotropy
 we move from the ferromagnetic $(\Delta=1)$ to 
the ferrimagnetic $(\Delta=-1)$ point. We
studied, by finite-size calculations and conformal invariance, four models of
 this kind; the $(S,S')$-mixed Heisenberg chains with $(S,S')=(\half,1)$ and
$(S,S')=(\half,\frac{3}{2})$, 
given in equation \rf{H} and the Heisenberg models with XXZ
interactions in the lattices of figures 2a and 2b ( models $ABC$ and $AB_2$ ).
We calculated the bulk and surface exponents of the first two models and the
bulk exponents of the last two.

All the models we studied show the universal feature of having a critical
phase for $1>\Delta>-1$ with long-distance physics governed by a $c=1$
Gaussian-like conformal field theory. The critical exponents, along this
phase, are model-dependent continuous functions of the anisotropy. 
The 
sound velocity and compactification radius of the Gaussian theory go to zero
at the isotropic ferromagnetic $(\Delta=1)$ and ferrimagnetic $(\Delta = -1)$ 
points. This reflects the fact that at both points we should expect the 
appearance of quadratic dispersion relations.
 We
strongly believe that this is the general scenario for arbitrary Heisenberg
chains showing ferrimagnetism instead of antiferromagnetism.

\begin{center}
{\bf Acknowledgments }
\end{center}

It is a pleasure to acknowledge profitable discussions with M. J. Martins 
and J. C. Xavier. We also thank M. T. Batchelor for a careful reading of 
our manuscript. This work was supported in part by Conselho Nacional de
Desenvolvimento Cient\'\i fico - CNPq - Brazil, and by Funda\cao de Amparo 
\`a Pesquisa do Estado de S\~ao Paulo - FAPESP - Brazil.
\newpage
\newpage
\begin{center}
\Large
Figure Captions
\normalsize
\end{center}
\vspace{0.7cm}

\hspace{-.65cm}Figure 1 - Schematic values of 
the lowest eigenenergy in a  sector with
magnetization $S_z$ of the Hamiltonian \rf{H}. The ground-state energy is
$E_0$ and $(S-S')\frac{L}{2}$ is an integer.
\vspace{1cm}

\hspace{-.65cm}Figure 2 - Lattices where the (a) $ABC$ 
and (b) $AB_2$ quantum chains are
defined. At the circles we have spin-$\half$ 
SU(2) operators and along the lines
the interactions are given by \rf{H2} ( XXZ type ). The Hamiltonian 
$AB_2$ is invariant under a local gauge transformation which 
independently interchanges the spin operators inside a rectangle in (b).
\newpage
\Large
\begin{center}
Table Captions
\end{center}
\normalsize
%\vspace{.25cm}

\hspace{-.65cm}Table 1 - Estimates, for some values
 of the anisotropy $\Delta$, of the
conformal anomaly of the $(S,S')$-Heisenberg chain \rf{H} for 
$(S,S') = (\half,1)$ and $(S,S') = (\half,\frac{3}{2})$.
\vspace{.2cm}

\hspace{-.65cm}Table 2 - Extrapolated values of 
the finite-size sequences \rf{xn1} for the
Hamiltonian \rf{H} with $(S,S') = (\half,1)$ and some values of the anisotropy
$\Delta$. The extrapolations in the third and fourth column are obtained
from the sequences $x_{2}^{\thalf,1}(L)/x_{1}^{\thalf,1}(L)$
and $x_{1}^{\thalf,1}(L)/x_{1}^{\thalf,\thalf}(L)$, respectively. The values
 of $x_{1}^{\thalf,1}$ in the fifth column are obtained from standard 
finite-size scaling (see equation \rf{gon}).
\vspace{.2cm}

\hspace{-.65cm}Table 3 - Extrapolated values of 
the finite-size sequences \rf{xn1} for the
Hamiltonian \rf{H} with $(S,S') = (\half,\frac{3}{2})$ and some values of the 
anisotropy $\Delta$. The extrapolations in the third and fourth column are
obtained from the sequences $x_{2}^{\thalf,\ttt}(L)/x_{1}^{\thalf,\ttt}(L)$
and $x_{1}^{\thalf,\ttt}(L)/x_{1}^{\thalf,\thalf}(L)$, respectively.
\vspace{.2cm}

\hspace{-.65cm}Table 4 - Dimensions $x_{1}^{S,S'}(\Phi)$ 
obtained from the bulk limit 
of the finite-sequences \rf{xPhiL} obtained from the Hamiltonian \rf{H}
with twisted boundary conditions (\ref{twis1bc},\ref{twis2bc}) specifyed
by the angle $\Phi = \pi$ and $\Phi = \frac{4\pi}{3}$. The values in 
parentheses are the predicted ones obtained by using the
values of $x_{1}^{S,S'}$ estimated in tables 2 and 3 in \rf{xPhi}.
\vspace{.2cm}

\hspace{-.65cm}Table 5 - Surface critical 
exponents $x_{s}^{S,S'}(n)$ associated to the
lowest eigenergies in the sectors $n=1,2$ of the Hamiltonian \rf{H}
with $(S,S') = (\half,1)$ and $(S,S') = (\half,\frac{3}{2})$ 
for some values of $\Delta$.
These estimates are obtained from the sequences \rf{xFree}.
\vspace{.2cm}

\hspace{-.65cm}Table 6 - Anomalous dimensions 
$x_n^{ABC}$ associated to the lowest eigenenergies
 in the sector $n=1,2$ of the model $ABC$ defined in figure 2b, for some
values of the anisotropy $\Delta$. The third and fourth column are calculated
similarly as for tables 2 and 3.
\newpage
%
% mudan@a do tamanho da pagina caso exista uma tabela muito larga
%
%\topmargin=1.7cm
%\textwidth=18.0cm
%\textheight=22.0cm
%\voffset=-3.2cm             
%\hoffset=-2.1cm
%
\Large
\begin{center}
Table 1
\normalsize
\vspace{1.cm}

\begin{tabular}{|c|c|c|c|c|c|}	\hline
$(S,S')		$	&$\Delta = .8090$ &$\Delta = 0.5$ &$\Delta = 0$&$\Delta = -0.5$&$\Delta = -0.7071$ \\ \hline
$(\half,1)	$	&0.93$\pm$0.05	&1.01$\pm$0.01	&1.03$\pm$0.04	&1.01$\pm$0.01	&0.9$\pm$0.1  \\
$(\half,\frac{3}{2})	$	&1.00$\pm$0.01  &0.99$\pm$0.01	&1.000$\pm$0.005&1.00$\pm$0.01	&0.9$\pm$0.1  \\ \hline
\end{tabular}
\vspace{2cm}

\Large
Table 2
\normalsize
\vspace{1cm}

\begin{tabular}{|c|c|c|c|c|c|}	\hline
$\Delta	$	&$x_{1}^{\thalf,1}$ &$x_{2}^{\thalf,1}$ &$x_{2}^{\thalf,1}/x_{1}^{\thalf,1}$&$x_{1}^{\thalf,1}/x_{1}^{\thalf,\thalf}$&$1-\gamma_{p}/2\nu$ \\ \hline
0.9172		&0.0414$\pm$0.0001	&0.165$\pm$0.001	&3.9996$\pm$0.0005	&0.664$\pm$0.005	&0.038  \\
0.80901		&0.0639$\pm$0.0005	&0.255$\pm$0.001	&3.9999$\pm$0.0005	&0.653$\pm$0.008	&0.069  \\
0.5		&0.1041$\pm$0.002	&0.415$\pm$0.001	&3.999$\pm$0.005	&0.626$\pm$0.001 	&0.104 \\
0.1736		&0.1302$\pm$0.0005	&0.521$\pm$0.004	&3.99$\pm$0.01		&0.584$\pm$0.002	&0.129 \\
0		&0.139$\pm$0.001	&0.554$\pm$0.002	&3.999$\pm$0.004	&0.554$\pm$0.002	&0.138  \\
-0.5		&0.136$\pm$0.001	&0.547$\pm$0.005	&4.01$\pm$0.01		&0.40$\pm$0.01		&0.134  \\
-0.7071		&0.105$\pm$0.005	&0.44$\pm$0.02		&4.01$\pm$0.01		&0.31$\pm$0.01		&0.111  \\
-0.9010		&0.035$\pm$0.005	&0.15$\pm$0.02		&3.99$\pm$0.02		&0.11$\pm$0.01		&0.066  \\ \hline
\end{tabular}
\newpage

\Large
Table 3
\normalsize
\vspace{.5cm}

\begin{tabular}{|c|c|c|c|c|}	\hline
$\Delta	$	&$x_{1}^{\thalf,\ttt}$ &$x_{2}^{\thalf,\ttt}$ &$x_{2}^{\thalf,\ttt}/x_{1}^{\thalf,\ttt}$&$x_{1}^{\thalf,\ttt}/x_{1}^{\thalf,\thalf}$ \\ \hline
0.9172		&0.0310$\pm$0.0005	&0.124$\pm$0.002	&3.9995$\pm$0.0005	&0.50$\pm$0.02		   \\
0.80901		&0.047$\pm$0.001	&0.189$\pm$0.002	&4.001$\pm$0.001	&0.47$\pm$0.02		  \\
0.5		&0.076$\pm$0.002	&0.301$\pm$0.002	&4.0000$\pm$0.0001	&0.44$\pm$0.01 		  \\
0.1736		&0.0914$\pm$0.0001	&0.3661$\pm$0.0002	&4.0001$\pm$0.0001	&0.412$\pm$0.001	  \\
0		&0.0959$\pm$0.0001	&0.384$\pm$0.001	&4.0000$\pm$0.0001	&0.383$\pm$0.001	  \\
-0.5		&0.089$\pm$0.002	&0.357$\pm$0.003	&4.00$\pm$0.04		&0.26$\pm$0.01		  \\
-0.7071		&0.071$\pm$0.002	&0.28$\pm$0.01		&3.98$\pm$0.01		&0.19$\pm$0.01		  \\
-0.9010		&0.031$\pm$0.001	&0.11$\pm$0.01		&3.99$\pm$0.02		&0.065$\pm$0.008	  \\ \hline
\end{tabular}
\vspace{1.cm}

\Large
Table 4
\normalsize
\vspace{.5cm}

\begin{tabular}{|c|c|c|c|c|c|}	\hline
					&$\Delta = 0.8090$ &$\Delta=0.1736$ &$\Delta=0$&$\Delta=-0.5$&$\Delta=-0.7071$ \\ \hline
$x_{1}^{\thalf,1}(\Phi=\pi)$		&0.93$\pm$0.02	&0.49$\pm$0.01	&0.46$\pm$0.01	&0.48$\pm$0.01	&0.58$\pm$0.01   \\
					&(0.978)	&(0.480)	&(0.4500	&(0.460)	&(0.595) 	\\
$x_{1}^{\thalf,1}(\Phi=4\pi/3)$		&1.65$\pm$0.05	&0.86$\pm$0.01	&0.81$\pm$0.01	&0.84$\pm$0.02	&1.03$\pm$0.02  \\
					&(1.74)		&(0.853)	&(0.799)	&(0.817)	&(1.06)		\\
$x_{1}^{\thalf,\ttt}(\Phi=\pi)$		&1.26$\pm$0.05	&0.686$\pm$0.002&0.657$\pm$0.005&0.70$\pm$0.01	&0.90$\pm$0.03  \\
					&(1.33)		&(0.684)	&(0.652)	&(0.702)	&(0.880)	\\
$x_{1}^{\thalf,\ttt}(\Phi=4\pi/3)$	&2.22$\pm$0.05	&1.222$\pm$0.004&1.165$\pm$0.005&1.25$\pm$0.01	&1.53$\pm$0.04  \\ 
					&(2.36)		&(1.216)	&(1.159)	&(1.248)	&(1.565)	\\ \hline
\end{tabular}
\newpage

\Large
Table 5
\normalsize
\vspace{1cm}

\begin{tabular}{|c|c|c|c|c|}	\hline
$\Delta	$	&$x_{1}^{\thalf,1}$ &$x_{2}^{\thalf,1}$ &$x_{1}^{\thalf,\ttt}$&$x_{2}^{\thalf,\ttt}$ \\ \hline
0.9172		&0.082$\pm$0.002	&0.330$\pm$0.005	&0.061$\pm$0.001	&0.245$\pm$0.005   \\
0.80901		&0.128$\pm$0.001	&0.513$\pm$0.003	&0.098$\pm$0.003	&0.392$\pm$0.005  \\
0.1736		&0.258$\pm$0.002	&1.036$\pm$0.003	&0.183$\pm$0.001	&0.73$\pm$0.01	  \\
0		&0.272$\pm$0.005	&1.106$\pm$0.003	&0.191$\pm$0.001	&0.766$\pm$0.005  \\
-0.5		&0.266$\pm$0.002	&1.070$\pm$0.005	&0.181$\pm$0.002	&0.728$\pm$0.005  \\
-0.9010		&0.1086$\pm$0.0005	&0.440$\pm$0.002	&0.065$\pm$0.005	&0.260$\pm$0.002 \\ \hline
\end{tabular}
\vspace{2cm}

\Large
Table 6
\normalsize
\vspace{1cm}

\begin{tabular}{|c|c|c|c|c|}	\hline
$\Delta	$	&$x_{1}^{ABC}$ 		&$x_{2}^{ABC}$ 		&$x_{2}^{ABC}/x_{1}^{ABC}$		&$x_{1}^{ABC}/x_{1}^{\thalf,\thalf}$ \\ \hline
0.9172		&0.07$\pm$0.01		&0.27$\pm$0.02		&3.99$\pm$0.02		&0.86$\pm$0.03	   \\
0.80901		&0.077$\pm$0.003	&0.31$\pm$0.01		&4.01$\pm$0.01		&0.84$\pm$0.02	  \\
0.5		&0.129$\pm$0.002	&0.51$\pm$0.01		&4.00$\pm$0.03		&0.78$\pm$0.02 	  \\
0.1736		&0.158$\pm$0.003	&0.635$\pm$0.005	&4.01$\pm$0.02		&0.71$\pm$0.01	  \\
0		&0.165$\pm$0.002	&0.667$\pm$0.005	&4.01$\pm$0.02		&0.66$\pm$0.01	 \\
-0.5		&0.152$\pm$0.002	&0.61$\pm$0.01		&3.99$\pm$0.02		&0.45$\pm$0.01	\\
-0.7071		&0.121$\pm$0.002	&0.486$\pm$0.002	&4.00$\pm$0.02		&0.32$\pm$0.01	 \\
-0.9010		&0.05$\pm$0.01		&0.20$\pm$0.01		&3.99$\pm$0.03		&0.11$\pm$0.01	  \\ \hline
\end{tabular}
\end{center}
\end{document}